\input harvmac
\def \la{\longrightarrow}

\def\mapa#1{\smash{\mathop{\longrightarrow }\limits_{#1} }}

\def\ass{$A_4^{sl(2)}$}

\def \del {\partial}

\def \ha  { {\textstyle{1\ov 2} } }

\def \a {\alpha}

\def\bs{\bigskip }

\def\us{s_{pq}}
\def\ut{t_{pq}}
\def\uu{u_{pq}}
\def\asl{A_{4}^{sl(2)}(s,t,u)}

\def\nn{\noindent }
\def\md{\medskip }
\def\slz{$SL(2,{\bf Z})$ }

\def \s {\sigma}
\def \p {\phi}

\def \n {\nu}

\def \td {\tilde }

\def \sm {\smallskip }

\def \inv {^{-1}}
\def \ov {\over }
\def \four{{\textstyle{1\over 4}}}

\def \lr { \lref}
\def\np {{  Nucl. Phys. }}
\def \pl {{  Phys. Lett. }}

\def \ijmp {{ Int. J. Mod. Phys. }}

\baselineskip8pt
\Title{
\vbox
{\baselineskip 6pt{\hbox{ }}{\hbox
{Imperial/TP/97-98/25}}{\hbox{hep-th/9802090}} {\hbox{
  }}} }
{\vbox{\centerline {Construction of $SL(2,Z)$ invariant  amplitudes}
\vskip4pt
 \centerline {in type IIB superstring theory}
}}
\vskip -27 true pt
\centerline  {  Jorge G. Russo{\footnote {$^*$} {e-mail address:
j.russo@ic.ac.uk
 } } 
}

\bigskip

 \centerline {\it  Theoretical Physics Group, Blackett Laboratory,}
\smallskip
\centerline {\it  Imperial College,  London SW7 2BZ, U.K. }

\medskip\bigskip

\centerline {\bf Abstract}
\medskip
\baselineskip10pt
\noindent
The construction of $SL(2,Z)$ invariant  amplitudes that generalize
the Virasoro amplitude is investigated in detail.
We describe a number of mathematical properties that characterize the simplest example, and present  pieces of evidence
that it represents the tree-level four-graviton
scattering amplitude
in membrane theory  on ${\bf R}^9\times T^2$
in the limit that the torus area goes to zero. 
In particular, we show that the poles of the $S$-dual amplitude are
 in precise correspondence with
the states of membrane theory that survive in the
type IIB limit. 
These are shown to be the states 
that span the Cartan subspaces of
area preserving diffeomorphisms of the 2-torus;
all other states  become infinitely massive,
and membrane world-volume theory acquires the structure
of a free theory.


\medskip
\Date {February 1998}
\noblackbox
\baselineskip 14pt plus 2pt minus 2pt

\lr\hult{
C.M. Hull and P.K. Townsend, Nucl. Phys. { B438} (1995) 109.}

\lr \green{M.B. Green and M. Gutperle, hep-th/9604091.}
\lr\mgreen{M.B. Green, hep-th/9712195.}

\lr \townelev{ P.K. Townsend, \pl B350 (1995) 184, hep-th/9501068.  }

\lr \witten {E. Witten, \np B460 (1995) 335.}

\lr\terras {A. Terras, {\it Harmonic Analysis on Symmetric Spaces and Applications}, vol. I, Springer-Verlag (1985).}

\lr \tow {P.K. Townsend,  hep-th/9609217.} 

\lr \schwa  {For reviews and further references see: J.H. Schwarz,
hep-th/9607201;
 M.J. Duff, hep-th/9608117;  P.K. Townsend,  hep-th/9609217.}
 
\lr \bergsh{ E. Bergshoeff, E.  Sezgin and P.K. Townsend, \pl B189 (1987)
75.}

\lr\banks{
T. Banks, W. Fischler, S.H. Shenker and L. Susskind, 
   hep-th/9610043.}
\lr \tser{A.A. Tseytlin, hep-th/9609212. }
\lr\russo {J.G. Russo, 
hep-th/9610018 .}
\lr \john{J.H. Schwarz, \pl B360 (1995) 13,
hep-th/9508143.}

\lr\rusty{J.G. Russo and A.A. Tseytlin, \np B490 (1997) 121.}

\lr\trusso{J.G. Russo, \pl B400 (1997) 37, hep-th/9701188.
}
\lr\mrusso{J.G. Russo, hep-th/9703118.}

\lr\jrusso{J.G. Russo, \pl B417 (1998) 253, hep-th/9707241.}

\lr\horst{G. Horowitz and A. Steif, \pl B258 (1991) 91.}

\lr\guven{R.~G\"uven, \pl B276 (1992) 49.}

\lr \gsh {M.B. Green  and J.H. Schwarz, \np B198 (1982) 441.}
\lr \gsb {
 M.B. Green, J.H. Schwarz and L. Brink,  \np B198 (1982) 474. }
\lr \mets { R.R. Metsaev and A.A.  Tseytlin, \np B298 (1988) 109. }
\lr \frats {E.S. Fradkin   and A.A. Tseytlin, \np B227 (1983) 252.}
\lr \duftom{ M.J. Duff and  D.J. Toms, 
in:  {\it Unification of the Fundamental 
Particle Interactions II},  Proceedings of the Europhysics Study Conference, 
Erice, 4-16 October 1981, 
ed. by J. Ellis and S. Ferrara (Plenum, 1983).} 
\lr \kallosh { R. Kallosh,  unpublished; P. Howe, unpublished. }
\lr \gris {M. Grisaru and W. Siegel,  \np B201 (1982) 292.}
\lr \sund {B. Sundborg,  \np B306 (1988) 545.}
\lr \acv { D. Amati, M. Ciafaloni and G. Veneziano, \ijmp  3 (1988) 615;
 \np B347 (1990) 550.}
\lr \lip { L.N. Lipatov, \pl B116 (1982) 411; \np B365 (1991) 614.}
\lr \cia {  M. Ademollo, A. Bellini and M. Ciafaloni, \np B393 (1993) 79.}
\lr \grisa{M.T. Grisaru, A.E.M. van de Ven and D. Zanon, \np B277 (1986) 388, 409.}
\lr\grow{D.J. Gross and E. Witten, \np B277 (1986) 1.}

\lr\creste{E.~Cremer, H.~Lu, C.N.~Pope and K.~Stelle, hep-th/9707207.}
\lr \bbpt{ K. Becker, M. Becker, J. Polchinski and A.A. Tseytlin, 
hep-th/9706072.   } 
\lr \sak { N. Sakai and Y. Tanii, \np B287 (1987) 457.}

\lr\polcha{J. Polchinski and  P. Pouliot, hep-th/9704029.}

\lr \witt { E. Witten, \np B443 (1995) 85, hep-th/9503124.}

\lr\kepa { A.~Kehagias and H.~Partouche, 
hep-th/9712164.}

\lr \ggv { M.B. Green, M.  Gutperle and P.  Vanhove, \pl B409
(1997) 177, hep-th/9706175. }

\lr \polch {J. Polchinski, {\it  TASI Lectures on D-branes},
hep-th/9611050.}
 
\lr \gv {M.B. Green and P. Vanhove, \pl B408 (1997) 122,
hep-th/9704145.}

\lr \grgu{M.B. Green and M. Gutperle, 
\np B498 (1997) 195, hep-th/9701093.}

\lr\tte{A.A. Tseytlin, \np B467 (1996) 383, hep-th/9512081.}
\lr\kiri{E. Kiritsis and B. Pioline, hep-th/9707018.}
\lr\rutse{J.G. Russo and A.A. Tseytlin, \np  {B508}
(1997) 245, hep-th/9707134.}
\lr\anto{I. Antoniadis, B. Pioline and T.R. Taylor, hep-th/9707222.}

\lr\anton{I. Antoniadis, E. Gava, K.S.~Narain and T.R.~Taylor, \np {B455} (1995) 109.}

\lr\towns{P.K. Townsend, hep-th/9705160; M.~Cederwall and
P.K.~Townsend, hep-th/9709002; N.~Berkovits, hep-th/9801009.}

\lr\bergsh{E. Bergshoeff,  E. Sezgin and P.K. Townsend, 
Ann. Phys. {185} (1988) 330.}

\lr\duf{M.J. Duff, P.S. Howe, T. Inami and K.S. Stelle, 
\pl B191 (1987) 70.}

\lr\bst{E. Bergshoeff,  E. Sezgin and Y. Tanii, \np {B298} (1988) 187.}

\lr\dewitt{B. de Wit, J. Hoppe and H. Nicolai, 
\np  {B305} [FS 23] (1988) 545.}

\lr \tte{A.A. Tseytlin, \np B467 (1996) 383, hep-th/9512081.}

\lr\ber{M. Bershadsky, S. Cecotti, H. Ooguri and C. Vafa, Commun. Math. Phys.
165 (1994) 31,  hep-th/9309140.}

\lr\bevaf{N. Berkovits and C. Vafa, hep-th/9803145. }

\lr\bvafa{N. Berkovits and C. Vafa, \np B433 (1995) 123;
H.~Ooguri and C.~Vafa, \np B451 (1995) 121.}

\lr\russo{For a recent review see: J.G. Russo, hep-th/9703118.}

\lr\berko{N. Berkovits, hep-th/9709116.}


\def \RR {{\cal R}}

\def\brs{\bar s}
\def\brt{\bar t}
\def\bru{\bar u}

\def\bs{\bigskip }

\newsec{Introduction}

With the premise that there is little prospect to determine an exact
scattering amplitude in M-theory, as in any non-trivial quantum theory
(especially if, as in the present case, the
 theory is  unknown), we  start by describing the 
purpose of this work.
The basic information about  four-graviton amplitude in
ten-dimensional type IIB superstring theory comes from three
different sources,
namely string perturbation theory,  low-energy results that 
are exact in the string coupling, and the symmetry under \slz 
transformations.
In addition, one must demand that 
any correction of perturbative origin  should appear
with an integer power of $g_B^2$ (type IIB string coupling),
and  non-perturbative corrections should be in 
 correspondence with D-instanton contributions.
Although these  ingredients are certainly insufficient to
anticipate the  general structure of the exact scattering
amplitude, simple examples that satisfy these
requirements can be constructed \jrusso . Here we will continue with this
program, and also 
illustrate how corrections can be systematically introduced
by preserving \slz symmetry at each step.\foot{
A possible reorganization of the perturbative expansion 
respecting \slz\ invariance was also suggested in ref.~\towns .}
In addition, we will find indications that
the simplest amplitude represents a special limit of
the tree-level diagram for the four-graviton amplitude
in membrane theory compactified on a 2-torus; this is the 
``type IIB" limit of M-theory on $T^2$, where
the torus area goes to zero at fixed moduli, so that
M-theory becomes the ten-dimensional type IIB superstring theory.

The \slz\ symmetry of  type IIB superstring theory \refs{\hult  \witten -\creste } requires that the effective action must be invariant under $SL(2,{\bf Z})$ transformations 
to all orders in the $\a' $ expansion.
 In the Einstein frame, a term of given order in derivatives 
involving the metric must be multiplied by a modular  function of the coupling. 
Since there is a one-to-one correspondence between certain (\slz invariant) terms in the effective action and the terms of the momentum expansion of an N-graviton amplitude, the same modular  functions appear in the N-graviton amplitude, which must therefore be invariant 
under $SL(2,{\bf Z})$ transformations. 

In section 2 we review the scattering amplitude proposed in ref.~\jrusso\ 
and, in addition, we describe a simple way to obtain it
by incorporating non-perturbative states of the spectrum.
The perturbative part 
of the S-dual scattering amplitude  can be resummed into a
simple closed expression, which is studied in section 3.
In section 4 the effective action that reproduces the
S-dual amplitude is examined.
In section 5 we show that this amplitude is uniquely determined
by a simple extra condition, the free wave equation in a locally flat three-dimensional space time parametrized by the 
type IIB string coupling $\tau=\tau_1+i\tau_2$ and the string tension.
Section 6 is an analysis of  more general \slz invariant amplitudes.
In  all cases, the (non-BPS) $(p,q)$ string states
of ten-dimensional type IIB theory play a central role.
The membrane configurations in eleven dimensions that give rise
to these states upon dimensional reduction and T-duality
are described in section 7, where we also make some remarks on 
a possible derivation
of the S-dual amplitude starting from eleven dimensions.
Finally, in section 8 we discuss the interpretation of the
results.

\newsec{Simplest \slz invariant amplitude}

\subsec{Definition and properties}

The scattering amplitude introduced 
in \jrusso\ is given by the following formula:
\eqn\vvzz{
A_4=\kappa^2  K \asl \ ,
}
\eqn\vvx{
\asl ={1\ov s t  u} \prod_{(p,q)'}
{\Gamma (1- \us)\Gamma (1-\ut)
\Gamma(1-\uu )\ov \Gamma (1+ \us )\Gamma (1+\ut )
\Gamma(1+\uu ) }\ ,
}
\eqn\stuu{
\us ={\a' s\ov 4|p+q\tau |}\ ,\ \ \ \ut ={\a' t\ov 4|p+q\tau |}\ ,\ \ \ 
\uu ={\a' u\ov 4|p+q\tau |}\ ,\ \ \ \ \us+\ut+\uu =0\ ,
}
where  $p$ and $q$ are relatively prime,
$\tau=C^{(0)}+i g_B^{-1}$ is the usual coupling of type IIB superstring theory, and
$K$ is the same kinematical factor depending on the momenta and 
polarization of the external
states appearing in the tree-level 
Virasoro amplitude of the form (see e.g. \grow )
$$
K=\zeta^{AA'}_1\zeta^{BB'}_2\zeta^{CC'}_3\zeta^{DD'}_4  K_{ABCD}(k_i)K_{A'B'C'D'}(k_i)\ ,\ 
$$
$$
 K_{ABCD}=-{1\ov 4}s t \ \eta_{AC}\eta_{BD}+...
$$
This amplitude can be obtained by the simple replacement
\eqn\ppr{
\sum _{m=1}^\infty {1\ov m^{2k+1}} \ \la\ \ha \sum_{(m,n)\neq (0,0)}
{1\ov |m+n\tau |^{2k+1}} \ ,
}
in the Virasoro amplitude,
\eqn\vrz{
A_4(s,t)=\kappa^2  K A_4^0(s,t) \ ,\ \ \ \ \ A_4^0 (s,t)={1\ov stu } e^{\delta _0(s,t)} \ ,
}
\eqn\logve{
\delta _0(s,t)= 2\sum_{k=1}^{\infty }
{\zeta (2k+1)\ov 2k+1} \big(\brs ^{2k+1}+\brt ^{2k+1}+\bru ^{2k+1} \big)\ ,
}
$$
\bar s=\four \a' s\ ,\  \ \ \bar t=\four \a' t\ , \ \ \ \bar u=\four
\a' u\ ,\ \ \ \ \brs+\brt+\bru =0\ ,
$$
The prescription \ppr\ is motivated by a number of facts:
\sm
\noindent a) it is the obvious generalization of the analogous replacement
$$
\sum _{m=1}^\infty {1\ov m^{3}} \ \la\ \ha \sum_{(m,n)\neq (0,0)}
{1\ov |m+n\tau |^{3}} \ ,
$$
in the first term of the sum \logve , which is known to account for
all perturbative and non-perturbative contributions to the $\RR^4 $ term
\refs{\grgu \gv -\berko }.

\sm

\noindent b) The same structure \ppr\ produces 
the non-perturbative contributions to the one-loop amplitude
coming from Kaluza-Klein gravitons (D0-branes) of $D=11$ 
supergravity \refs{\ggv , \rutse }. 
The D0-branes are related by duality to the D-instantons that are the origin of the non-perturbative effects in the ten-dimensional type IIB theory.

\sm

\noindent c) This prescription leads to correct
perturbative $g_B^{2k}$ and non-perturbative $O(e^{-2\pi m n/g_B})$
dependence, with $k,m,n$ integer numbers. 
This is non-trivial, and it is crucial
in order to have a one-to-one correspondence between these terms
and instanton contributions. [For example, an ansatz giving rise to $O(e^{-4\pi m n/g_B})$ dependence, could not be correct,
since it would miss some D-instanton configurations.] This property follows
by first writing  \ass\ in terms of Eisenstein series,
\eqn\ess{
E_r(\tau )=\sum_{(p,q)'} {\tau_2^r\ov |p+q\tau |^{2r} }\ ,
}
\eqn\unaa{
\asl = {1\ov st u} e^{\delta (s,t)}\ ,
}
\eqn\ddelta{
\delta (s,t) = 2 \sum _{k=1}^\infty {\zeta (2k+1) g_B^{k+1/2}
E_{k+1/2}(\tau )\ov 2k+1} 
\big(\brs ^{2k+1}+\brt ^{2k+1}+\bru ^{2k+1} \big)\ ,
}
and then using the expansion 
at large $\tau _2 $,
\eqn\bbb{
E_r(\tau )=\tau_2^r+\gamma_r \tau_2^{1-r}+
{4\tau_2^{1/2}\pi^r\ov\zeta(2r)\Gamma(r)}
\sum_{n,w=1}^\infty \big({w\ov n}\big)^{r-1/2}\cos(2\pi  wn\tau_1)K_{r-1/2}(2\pi w n\tau_2 )\ ,
}
$$
\gamma_r={\sqrt{\pi }\ \Gamma(r-1/2)\ 
\zeta(2r-1)\ov \Gamma(r)\ \zeta(2 r) }\ .
$$
Using the asymptotic expansion for the Bessel function $K_{r-1/2}$,
$$
K_{r-1/2}(2\pi w n\tau_2 )={1\ov \sqrt{4wn\tau_2 } }e^{-2\pi w n\tau_2}\sum_{m=0}^\infty {1\ov (4\pi wn \tau_2)^m }{\Gamma(r+m)\ov \Gamma(r-m)m! }\ ,
$$
we see that the $E_{k+1/2}(\tau )$ terms in the amplitude
are of the form
\eqn\qqq{
g_B^{k+1/2} E_{k+1/2}(\tau )=1+\gamma _{k+1/2}\ g^{2k}_B
+O\big( e^{-2\pi/g_B}\big)\ .
}

\sm

\noindent d) It gives an $SL(2,{\bf Z})$ invariant amplitude
that in the limit $g_B^2\to 0$ reduces to the Virasoro amplitude
($SL(2,{\bf Z})$ invariance is explicit in the Einstein frame,
$g_{\mu\nu}^E=g_B^{-1/2} g_{\mu\nu}$, so that $s_E=g_B^{1/2} s$, etc.).

\sm

\noindent e) The resulting amplitude has poles in the $s$-$t$-$u$ channels at $\us=-n $, $\ut=-n $,
$\uu=-n $, $n=0,1,2,...$
corresponding to exchange of particles with masses
\eqn\polos{
\four \a' M^2 =n |p+q\tau |\ ,
}
which is precisely the desired  spectrum of  $(p,q)$ string states:
\eqn\ppqq{
 M^2=4\pi T_{pq}(N_R+N_L)={2\ov\a'} \ |p+q\tau |\ (N_R+N_L)\ ,\ \ \ \ N_R=N_L\ .
}
This spectrum corresponds to the zero winding sector of the spectrum
originally studied in  \john\ for the nine-dimensional type IIB
string theory.

\subsec{ \slz symmetric expressions  by including $(p,q)$ strings}

We start  with the Virasoro amplitude, with $\delta_0 $ as given in 
eq.~\logve . By writing the $\zeta $-functions as series, it takes the form
\eqn\vvxx{
\delta_0 =2 \sum_{k=1}^\infty \sum_{m=1}^\infty {1\ov 2k+1} 
\left({\a 's\ov 4m}\right)^{2k+1}
\  +(s\to t)+ (s\to u)
}
or
\eqn\suit{
\delta_0 = \sum_{m=1}^\infty \delta _{(m)} \ ,
\ \ \ \delta _{(m)}= \log {M_m^2+ s\ov M_m^2- s}
\  +(s\to t)+ (s\to u)
\ , \ \ \ \a' M_m^2=4m \ .
}
Thus $\delta_0 $ is a direct sum of $\delta _{(m)} $ associated
with each mass level.
This form is suitable for \slz symmetrization: we just need 
to include in the sum \suit\ all contributions $M^2_{mn}$
associated 
with the masses of $(p,q)$ string states,
\eqn\esta{
\delta  = \ha \sum_{(m,n)\neq (0,0)}  \log {M^2_{mn}+ s\ov M^2_{mn}- s}
\  +(s\to t)+ (s\to u)
}
i.e.
$$
\delta =\ha \sum_{(m,n)\neq (0,0)} \log 
{ 4|m+n\tau | + \a' s\ov 4|m+n\tau | -\a' s}\  +(s\to t)+ (s\to u)
\ . 
$$
Expanding the logarithm, this becomes
\eqn\exxx{
\delta=  \sum_{(m,n)\neq (0,0)} \sum_{k=1}^{\infty }
{1 \ov 2k+1}\left( {\a' s\ov 
4|m+n\tau|}\right)^{2k+1} \  +(s\to t)+ (s\to u)
}
$$
=2 \sum_{(p,q)'}\sum_{k=1}^\infty 
{\zeta (2k+1) \ov 2k+1}\big( \us ^{2k+1}+
\ut ^{2k+1}+\uu ^{2k+1}\big) \ ,
$$
where $\us, \ut, \uu $ were introduced in eq.~\stuu .
It is perhaps  not a surprise that we obtain just the same amplitude
\ass \ \foot{
There is a curious representation of the 
Virasoro amplitude in terms  of free fermion 
variables $d_m, d^{\dagger }_m$. 
Using $
\log{b\ov a}= \int_0^\infty {dt'\ov t'} \big( e^{-a t'}
-e^{-b t'} \big) \  ,
$
one can write 
$$
\delta_0  = -
\int_0^\infty {dt'\ov t'} {\rm Tr}\big[ (-1)^F e^{-t'H }\big]\ .
$$
 where 
$ H=- \four \a' p^2 + 2 \sum_{m=1}^\infty m (d^{\dagger }_m d_m -\ha )
\ ,\ \  F=\sum_{m=1}^\infty d^{\dagger }_m d_m\ .
$ 
It is clear that  $\delta $ defining $A_4^{sl(2)}$ 
is given  by the same expression with 
$
H=-\four \a' p^2 +  \sum \omega_{mn} (d^{\dagger }_{mn} d_{mn} -\ha )
 ,\   \omega_{mn}=|m+n\tau |\ .
$
The frequency $\omega_{mn}$ can be associated with the
oscillations of membranes that move along a 
$(p,q)$ cycle  
of the 2-torus carrying zero total momentum 
(with $p/q=m/n$, see sect. 7). 
}
\eqn\masmas{
\asl ={1\ov stu} e^{\delta (s,t,u)} \ .
}
Along with the fundamental property described in    section 5, 
this shows that  simple attempts of \slz symmetrization 
indeed lead to the amplitude \ass . One might also attempt
to construct an \slz invariant amplitude by
replacing the product over $(p,q)'$ states in eq.~\vvx\ 
by a sum over $(p,q)'$  states.
This does not lead to a sensible amplitude: 
first of all, the leading term in the expansion in $\a' $ is divergent;
in addition, such object contains perturbative dependence in odd
powers of $g_B$, which cannot arise in string theory.
Let us also point out that obtaining \slz symmetric terms in the effective action
by summing over $(p,q)$ was  recently  investigated in \kepa .

\newsec{Resummation of perturbative part }
As shown in \jrusso , and it is clear from eqs. \unaa -\bbb ,  
the amplitude \ass\ can be written as
\eqn\vvv{
 A_{4}^{ sl(2)}(s,t,u)= A_{4}^{ \rm pert}(s,t,u)
\ +\ O\big( e^{-2\pi/g_B}\big)\ ,
}
\eqn\dos{
  \ A_{4}^{\rm pert }(s,t,u)=A_4^0(s,t,u) \ e^{ \td A_4(s,t,u)}\ ,
}
with
\eqn\fff{
 \td A_4(s,t,u)=\sqrt{\pi } \sum_{k=1}^\infty
{(k-1)!\zeta (2 k)\ov \Gamma(k+3/2) }g_B^{2k}  
\big(\brs ^{2k+1}+\brt ^{2k+1}+\bru ^{2k+1} \big)\ .
}
Writting $\zeta (2k)=\sum_{m=1}^\infty m^{-2k } $,
and using the formula
$$
{\sqrt{\pi} \ov 4}\ \sum_{k=1}^\infty {(k-1)!\ov 
\Gamma (k+3/2)} x^k =1-\sqrt{ {1\ov x}-1 } \ {\rm arcsin}\sqrt{x}\ ,\
\ \ |x|<1\ ,
$$
the perturbative part of \ass\ can be resummed
with the result
\eqn\rssm{
\td A_4(s,t,u)=-4 \sum_{m=1}^\infty \sqrt{ {m^2\ov g_B^2}-\brs ^2 } \ {\rm arcsin}{\brs 
g_B\ov m}
\ + (s\to t)+ (s\to u)\ ,
}
or
\eqn\resu{
A_{4}^{\rm pert }(s,t,u)=A_4^0(s,t,u) \prod _{m=1}^\infty
\left[ {i\brs g_B\ov m} +\sqrt{1-{\brs ^2g^2_B\ov m^2}}\right]^
{ 4i\sqrt{{m^2\ov g_B^2}-\brs ^2 } }\times [s\to t]\times [s\to u]\ ,
}
where we have written ${\rm arcsin}{z} =-i\log \big(
iz+\sqrt{1-z^2}\big)$.

Although eq.~\resu\ may suggest the presence of cuts for 
$\brs >g_B\inv$, the amplitude $A_{4}^{\rm pert }$
cannot be extended to this regime; the terms
$ O\big( e^{-2\pi/g_B}\big)\ $ that so far have been neglected 
become important.
The full expression \vvx\ indicates that \ass\ has no cuts.
This can be proved with no need of 
  understanding 
the convergence properties of the infinite product in \vvx .  
Indeed, the presence of a cut at $\a' s_E > 4/g_B^{1/2}$ would
imply, by S-duality, the presence of a cut at $\a' s_E > 4 g_B^{1/2}$.
For sufficiently small $g_B$, and $s_E,t_E,u_E $ fixed, eq.~\resu\
is applicable at $\a' s_E > 4 g_B^{1/2}$ (viz. $\brs =1$), 
and it has no discontinuity cut at that point. 
Therefore there cannot be any cut at any \slz rotation of
this condition, in particular, at $\a' s_E > 4/g_B^{1/2}$.

The original expression \fff\ already exhibits the fact 
that there is no absorption via opening
of inelastic channels, since for sufficiently small $g_B$ it is convergent
at $\bar s=1$ and real;  the above argument can then be applied for any 
\slz rotation of this point. It is interesting to note that
in the physical region of the parameter space one has
$$
s>0\ , \ \  \ \ 0< -{t\ov s}=\sin^2 {\p \ov 2} < 1\ ,
\ \ \ \ 0< -{u\ov s}=\cos^2 {\p \ov 2} < 1\ ,
$$
and similar conditions in the regions $t>0$ or $u>0$. Hence
$$
\brs ^{2k+1}+\brt ^{2k+1}+\bru ^{2k+1}=\brs^{2k+1}
\big[1- (\sin^2 {\p \ov 2} )^{2k+1} - 
(\cos^2 {\p \ov 2} )^{2k+1}\big] >0 \ .
$$
Thus (in the region the sum converges, $g_B\brs <1 $)
 $\td A_4$  is a real number greater than zero, so that
$$
\left| A_{4}^{\rm pert }(s,t,u)\right|
=\left| A_4^0(s,t,u)\right| \ e^{ \td A_4(s,t,u)} \  >\ 
\left| A_4^0(s,t,u)\right|
$$
In particular, this also indicates that including the 
contribution of $(p,q)$ states
increases the probability amplitude of the process.
For $\a' s > 4/g_B$, this analysis is not applicable, 
and the general expression \vvx\ must be used.

\newsec{Effective action }

We would like to explore the structure of the effective action that reproduces \ass .
It is convenient to introduce the notation:
\eqn\fzzz{
f_k(\tau )={\zeta (2k+1)\ov k+\ha }\ g_B^{k+\ha } E_{k+1/2 }(\tau )\ ,
}
in terms of which $\delta $ (defined in \ddelta ) becomes ($\a '=4 $)
\eqn\dfz{
\delta = \sum_{k=1}^\infty \ f_k(\tau ) \ (s^{2k+1}+t^{2k+1} +u^{2k+1} ) \ .
}
Expanding $A_4^{sl(2)}(s,t) $ in 
eq.~\unaa\ in powers of $\delta $, the amplitude \vvzz\ 
exhibits the pole due to the exchange of the massless
supergravity multiplet plus and infinite number of terms containing
polynomials in $s,t$, 
\eqn\exz{
A_4(s,t)=\kappa^2 K \bigg[ {1\ov st u} 
+ \sum_{k=1}^\infty f_k(\tau ) P_k(s,t)
+...
+{1\ov N!}\sum_{k_1...k_N} f_{k_1}(\tau )... f_{k_N} (\tau ) P_{k_1....k_N}(s,t)+...\bigg]\ ,
}
with
\eqn\ppz{
P_k(s,t)= {1\ov stu } (s^{2k+1}+t^{2k+1}+u^{2k+1})\ ,
}
\eqn\qqz{
P_{k_1...k_N}(s,t)=P_{k_1}...P_{k_N} (stu)^{N-1} \ .
}
Note that $P_k(s,t)$ is an homogeneous polynomial of degree $2k-2$ in $s,t$,
as it is clear after using $u=-s-t $,
\eqn\ppzz{
P_k(s,t)=
\sum_{l=1}^k {(2k+1)!\ov l! (2k+1-l)!} \sum_{n=0}^{2k-2l} (-1)^n s^{2k-1-l-n} t^{l+n-1}  \ .
}

At $g_B\ll 1$, $f_k(\tau )$ has the expansion (see eq.~\qqq )
\eqn\efz{
f_k(\tau )={\zeta (2k+1)\ov k+\ha } \ +\ {\sqrt{\pi }\Gamma (k)\zeta (2k)\ov 
\Gamma (k+ {\textstyle{3\ov 2} } ) }\ g_B^{2k}\ +\ 
O(e^{-2\pi/g_B} )
}
Using eq.~\efz , we see that at weak coupling the 
generic $N$ term in eq.~\exz\ takes
the form 
\eqn\esof{
{1\ov N!stu}\delta^N= \sum_{k_1...k_N} \big[ c_1+c_2 g_B^2+...+c_h g_B^{2h}+
O(e^{-2\pi/g_B}) \big] P_{k_1...k_N}(s,t) \ ,
}
$$
h=2(k_1+...+k_N)\ .
$$
The first term $N=1$ has only tree-level and genus $k$ contributions, and corresponds to  local terms in the effective action of the form
$$
S_{A_4}\bigg|_{N=1}
=\int d^{10}x \sqrt{-G}\ g_B^{-2}\ 
\sum_{k=1}^\infty \ f_k(\tau ) \ ``\nabla ^{4k-4} \RR^4 "\  
$$
\eqn\effz{
=\int d^{10}x \sqrt{-G}\ g_B^{-2}\ \sum_{k=1}^\infty \ f_k(\tau )\ P_k (s,t) \RR^4\  
}
$P_{k_1...k_N}(s,t)$ is an homogeneous polynomial in $s,t $ of degree $2h+N-3$.
Therefore, there are also contributions to the order $``\nabla ^{4k-4} \RR^4 " $ in derivatives 
coming from terms with $N>1$, $N$ odd, $2h=2k+1-N$. 
It is clear from \ppz, \qqz \ that the tensor structure of 
each of such terms is different; they give new terms to the effective action which do not mix with \effz .

 {}From eq.~\esof\ we see that the
perturbative contribution in a generic
$N$ term $``\nabla ^{4k-4} \RR^4 " $  of highest order  corresponds to genus 
$k - \ha (N-1)\leq k$.
Similarly, terms with $N$ even contribute to $``\nabla ^{4k-2} \RR^4 " \ ,\ k=2,3,...$, with  perturbative contribution of highest order corresponding to genus
$k-\ha (N-2)\leq k$. Thus either $``\nabla ^{4k-4} \RR^4 " $ or 
$``\nabla ^{4k-2} \RR^4 " $ do not receive contributions beyond genus $k$.

In a recent paper, Berkovits and Vafa \bevaf \ conjectured that 
the exact function of the coupling 
for the term $H^{4k-4} \RR^4 $ in the type IIB effective action on ${\bf R}^{10}$  is given by $ f_k(\tau )$
(up to a numerical multiplicative constant).
This  was based on explicit genus $k$ results
\bvafa\  (see also  \refs{\anton ,\ber } for analogous  results in four dimensions), and it is also supported
by the fact that such terms can only receive tree-level and genus $k$ contributions.
As observed in \bevaf , this conjecture should be  related to   
the  amplitude \ass\ conjectured 
in \jrusso\ by virtue of supersymmetry.
 By a slight elaboration of the argument of ref.~\bevaf , 
let us now argue that the $H^{4k-4} \RR^4 $   conjecture 
actually {\it implies} that the terms containing $P_k(s,t)\RR^4 $ in the {\it exact} four-graviton amplitude must be multiplied by $f_k(\tau )$.
Supersymmetry transformations are expected to 
relate the term $H^{4k-4} \RR^4 $   to 
a term with no $H$ field but with the same number of derivatives,
i.e. of the form $``\nabla ^{4k-4} \RR^4 " $. The tree-level contribution to 
$H^{4k-4} \RR^4 $   contains a single factor $\zeta (2k+1) $. Because supersymmetry transformations
on a given term cannot generate  $\zeta (2k+1)$ factors 
(there are no zeta-functions in the supersymmetry transformation laws),
the term $``\nabla ^{4k-4} \RR^4 " $ that is in the same supersymmetric
invariant as $H^{4k-4} \RR^4 $   must also contain a single factor  $\zeta (2k+1)$.
There is only one tree-level term of order  $``\nabla ^{4k-4} \RR^4 " $ that contains
 such single factor, namely $P_k(s,t) \RR^4 $, which must therefore be in the same supersymmetric invariant as $H^{4k-4} \RR^4 $.
Thus,  the exact function of the coupling multiplying 
 $P_k(s,t) \RR^4 $ must be proportional to the exact 
function of the coupling   multiplying $H^{4k-4} \RR^4 $.   

In this way one  {\it derives} eq.~\effz\ 
from the $H^{4k-4} \RR^4 $   conjecture of \bevaf .
It is remarkable that these two completely independent approaches
have led to the same result. Unexpectedly, this holds for terms of arbitrarily high orders in derivatives.
If the conjecture of \bevaf\ is true,  then the {\it exact} four-graviton 
amplitude can  differ from \ass\  only in extra non-local pieces (which are certainly expected),
and   in higher genus contributions to the terms $P_{k_1...k_N}(s,t)$ with $N>1$. Possible corrections
to the local terms with $N>1$ are however highly constrained by the fact
that they must not give any tree-level 
contribution and they must be $SL(2,{\bf Z})$ invariant by themselves,   in addition to having correct perturbative $g^{2k}_B$ and non-perturbative $O(e^{-2\pi mn/g_B})$ dependence on the coupling.
It should be noted that the terms $P_{k_1...k_N}(s,t)\RR^4 $ with $N>1$ cannot be related by supersymmetry
to $H^{4k-4} \RR^4 $ for the reasons explained above, i.e.  the ratio of the
coefficients of the respective tree-level parts is irrational. Thus the knowledge of
terms $H^{4k-4} \RR^4 $ does not help in determining the exact functions of the coupling
that multiplies $P_{k_1...k_N}(s,t)\RR^4 $ terms with $N>1$.

 The function of the coupling in front of each term $(\nabla^2)^{n}\RR^4$ will in general 
 be given by a product of Eisenstein functions.
In particular, the first terms $\RR^4 ,\  \nabla^{4}\RR^4 , \
\nabla^6\RR^4$ in the effective action are of the form
$$
S'=\int d^{10}x \sqrt{-G_E}\bigg(2\zeta(3) E_{3/2}(\tau )
\RR^4_E +
2\zeta(5) E_{5/2}(\tau ) \nabla ^4 \RR^4_E
$$
\eqn\ayay{
+\ 2 \zeta(3) ^2 E_{3/2}^2(\tau )\nabla ^6\RR^4 _E \bigg)\ .
}
In the above notation, these terms correspond to $P_1(s,t), \ P_2(s,t),\ P_{11}(s,t)$, respectively.
In the string frame, this takes the following
form at $g_B^2\ll 1$
$$
S'=\int d^{10}x \sqrt{-G}\bigg(
(a_0 g_B^{-2}+a_1) \RR^4 
+(b_0 g^{-2}_B + b_2 g^2_B ) 
\nabla ^4 \RR^4 
$$
\eqn\ggbb{
+(c_0 g^{-2}_B+c_1+c_2 g^2_B )\nabla ^6\RR^4 \bigg)+O(e^{-2\pi /g_B})\ .
}
We observe that there is no genus-one contribution to the order $s^2$
($\sim \nabla ^4 \RR^4 $).
The genus-one contribution to $s^3$ is given by
$4\zeta(3) \zeta (2)= {2\pi^2\ov 3}\zeta (3) $.

\newsec{Fundamental property of $\asl $}

\def\ff{${\cal F}$ }

Another strategy to produce an \slz invariant amplitude is by
generalizing, when possible, properties of the Virasoro amplitude 
(which applies at $\tau_2\to \infty $) to the
full fundamental domain \ff $=SL(2,{\bf Z})\setminus H$.
Let us write the Virasoro amplitude \logve\ 
in terms of Einstein variables, $s_E=g_B^{1/2} s$, etc.

\eqn\eins{
\delta_0 (\tau_2 ,\a')=  2 \sum_{k=1}^\infty 
c_k(s_E,t_E) \tau_2^{k+1/2}(\a') ^{2k+1}\ ,
}
with
$$
c_k(s_E,t_E)={\zeta (2k+1)\ov 2k+1}\big(
s_E^{2k+1}+t_E^{2k+1}+u_E^{2k+1} \big)\ .
$$
As a function of $\tau $ and $\a' $, $\delta_0 $ obeys the simple relation
%
$$
\tau_2^2 {\del ^2 \ov \del \tau_2^2}\delta_0=
\eta^2 {\del ^2 \ov \del \eta^2}\delta_0\ \ ,\ \ \ \ 
\eta\equiv (\a')^2\ .
$$
The \slz invariant condition is thus the following one:
\eqn\lll{
\Delta \delta_0= \eta^2 {\del ^2 \ov \del \eta ^2}\delta_0\ ,
}
where $\Delta $ is the Laplace operator in the fundamental domain, 
$$ 
\Delta= \tau_2^2 
\big({\del^2\ov \del \tau_1^2}+ {\del^2\ov \del \tau_2^2}\big)\ .
$$
Any solution to eq.~\lll\ in \ff that asymptotically approaches 
$\delta_0 $ will provide an \slz generalization of the 
Virasoro amplitude. Interestingly, 
this strategy leads to the same amplitude discussed
in sect.~2, as stated by the following theorem.
\md

\noindent {\it Theorem:} 
Let $\delta (\tau ;\eta )$ be a function on \ff satisfying
\eqn\thth{
 \lim _{\tau_2\to \infty }\delta(\tau ;\eta )=\delta_0\ ,\ \ \
}
\eqn\hhh{  
\Delta \delta = {\eta }^2 {\del^2\ov \del {\eta }^2}  \delta\ ,\ \ \ \ 
\eta=(\a')^2\ ,
}
where the limit \thth\ is understood with $\eta\to 0$ so that
 $\tau_2\eta $ is fixed (``string frame").
Then such function is unique and given by 
$$
{1\ov stu} e^{\delta }=\asl \ .
$$

\noindent {\it Proof:} We solve eq.~\hhh\ by separation of variables: 
\eqn\sss{
\delta= \sum_r f_r(\eta )\psi_r (\tau )\ .
}
It follows that
\eqn\ssz{ 
f_r(\eta )= A_r \eta^r +B_r \eta ^{1-r}\ ,
}
\eqn\kkk{
\Delta \psi_r= r(r-1) \psi _r\ . 
}
Since $\psi_{1-r}$ lies in the space of solutions $\psi_r $ to \kkk ,
with no loss of generality we can set $B_r=0$.
Using eq.~\thth , we find that  only $\psi _r $ with $r=k+1/2$
appear in the sum \sss , and they have the asymptotic behavior
$$
\psi_r \ \mapa{\tau_2 \to \infty }\ \delta_{r,k+{1\ov 2} }\tau_2^r\ ,\ \ \ \ k=1,2,... 
$$
This implies that $\psi_r(\tau )$ is a Maass waveform.
[A Maass waveform is a function on ${\cal F}$ which is an eigenfunction
of the laplacian and which has at most polynomial growth at infinity
\terras .]
If ${\cal N}(SL(2,{\bf Z}),r(r-1))$ denotes the vector space of such 
waveforms, it is a known result that 
$$
{\cal N}(SL(2,{\bf Z}),r(r-1))={\bf C} E_r\ ,\ \ \ \ {\rm for}\ {\rm Re}\ r>\ha \ ,
\ \ r\not\in [\ha ,1] \ .
$$
Indeed, if there was another function $f$ with the same asymptotic behavior,
it would imply that we can find a constant $c$ such that $g=f-c E_r$
is square-integrable over the fundamental domain, with the
invariant area element $d^2\tau /\tau^{2}_2$ (since its expansion
would start with $b \tau_2^{1-r}$, see eq.~\bbb). But this contradicts the fact that the Laplace operator is negative on \ff \terras. 

Using the boundary condition, we can now determine the coefficients $A_r$:
\eqn\ffg{
\delta =2 \sum_{k=1}^\infty c_k(s_E,t_E) E_{k+1/2}(\tau )\eta ^{k+1/2} \ ,
}
or
\eqn\ffz{
{1\ov stu} e^{\delta }=\asl \ ,
}
Q.E.D.

\medskip

Thus the differential equation \hhh\ can be used to provide an 
alternative definition  for $\asl $, with no need to
refer to the  infinite product  \vvx , whose convergence
properties are unknown. 
What is special about this differential equation? 
In the three-dimensional ``space-time" with coordinates 
$\{ \eta ,\tau_1, \tau_2 \}$ it takes the simple form
\eqn\wavv{
 \Delta_{(3)} \delta =0\ ,
}
with
\eqn\mtt{
ds^2_3=-\eta^{-4}d\eta^2 + {1\ov \tau_2^2 \eta^2}(d\tau_1^2+d\tau_2^2)\ ,
}
i.e.
\eqn\mttt{
ds^2_3=-d\tau_0^2+{\tau_0^2\ov \tau_2^2} (d\tau_1^2+d\tau_2^2)\ ,\ \ \ \ 
\tau_0=\eta^{-1}=(\a ')^{-2} \ .
}
This geometry may be interpreted as
an ``expanding universe", with the spatial
section being the fundamental domain \ff .
It does not have an Euclidean counterpart.
The time parameter, which provides the scale, is the string tension
squared, $\tau_0=(2\pi T)^2$.

Introducing $U=\tau_0/\tau_2$\ , $V=\tau_0\tau_2$, we obtain
\eqn\zaz{
ds^2_3=-dUdV+ U^2 d\tau_1^2\ .\ 
}
In this form, the geometry exhibits an orbifold singularity at $U=0$
moving at the speed of light.
This metric (with the range of $U,V$ unrestricted)
was called the  ``null orbifold" geometry 
in  \horst , but the connection with the fundamental
domain of \slz was not noticed.
Here the topology is not the same, since
there is a restriction in the range of $V/U$, 
and the geometry contains 
 singularities at $|\tau |=1, \tau_1=\pm 1/2$.
A new change of coordinate shows that the geometry is flat
everywhere away from the orbifold points,
\eqn\flt{
ds^2_3=-dUd\td V +dy^2\ ,\ \ \ \ \ y=U\tau_1\ ,\ \ \td V=V+U\tau_1^2\ .
}
In other words, this three-dimensional space-time is nothing but the
(Minkowskian) embedding of the fundamental domain.
The differential equation that defines the amplitude is thus 
the simplest
invariant differential equation that one can write down involving $\tau ,\a'$, namely the free wave equation in a flat three-dimensional space.
This differential equation involves R-R and dilaton couplings, and  $\a '$,
i.e. the length scale of the target metric.
It might originate from a Ward identity (or perhaps from some saddle point 
approximation to the scattering problem).


\newsec{More general $S$ dual amplitudes}

\def\aex{A_4^S (s,t)}

Since the exact amplitude is a function on the fundamental domain 
${\cal F}=SL(2,{\bf Z})\setminus H$, any correction to \ass\ must be invariant under
\slz transformations.
For square-integrable functions on \ff\ 
(with the standard measure $d\tau^2/\tau_2^2$~) there
exists a  spectral decomposition in terms of cusp forms and
$E_{1/2+i \phi }$, namely the Roelcke-Selberg formula  \terras .
Although it is not completely clear 
whether the exact amplitude must be in
$L^2(SL(2,{\bf Z})\setminus H)$,
it is interesting to note that a function
with the asymptotic behavior of the Virasoro amplitude would be square-integrable, since
the relevant asymptotic region is $\tau_2\to \infty$
at fixed $s_E,t_E, u_E$, in other words, $s,t,u\to \infty $,
where one has the well-known exponential fall off
of the  high-energy fixed-angle limit
(recall  
$(s,t,u)=\tau^{1/2}_2 (s_E,t_E,u_E) $). However, 
the exact scattering amplitude
must approach the Virasoro amplitude only at  $\tau_2\to \infty$
and fixed string-frame variables $s,t,u$. Thus,
in the region of interest, $\tau_2\to \infty$
at fixed $s_E,t_E, u_E$, the
asymptotic behavior of the exact amplitude is unknown.
This is a region of high-energy scattering at fixed angles,
where, in addition, the coupling $g_B$ is sent to zero.
It seems
legitimate to make use of some version of unitarity bounds, 
which could indicate that the exact amplitude
must be square-integrable 
(as far as square-integrability is concerned, even a 
field-theoretic behavior may be sufficient; in local field theory, 
high-energy fixed angle scattering amplitudes typically 
fall off according to a power law, 
fact directly associated with the power singularities of the products
of local operators at short distances). 
In any case,
it is unlikely that a treatment based on the
Roelcke-Selberg expansion would be of any use, 
since  an orthonormal
basis of cusp forms for \slz is not explicitly known.

The general structure of the exact scattering amplitude can be quite
complicated, but here we will consider a   subclass
of possible corrections.
They exhibit an interesting  feature:
each new order is \slz invariant by itself and, when expanded at $g_B\ll 1$, starts with an additional power
of $g_B^2$.
In addition, it provides an example of a more general \slz invariant amplitude that also reproduces  the Virasoro amplitude
in the weak coupling limit and has a correct perturbative and non-perturbative dependence on the coupling.


In the one-loop four-graviton amplitude of eleven-dimensional
supergravity on the torus \rutse , as well
as in \ass , only those $E_r(\tau )$, with $r=k+1/2=3/2,5/2,... $  
appear. Other $E_{r}$ are simply not allowed,
because they would contain wrong  
perturbative dependence at small $g_B$ (see \bbb ).
It is therefore of interest to investigate possible
 additional corrections to $\delta $ that can be
expressed as a linear combination of $E_{k+1/2}(\tau )$.
Let us stress once again that this does not represent the most 
general function on \ff . 
We shall thus consider  a function on ${\cal F}$ of the form
\eqn\amex{
\log \aex \equiv -\log stu + 2 \sum_{k=1}^\infty g_B^{k+1/2}
(\brs ^{2k+1} +\brt ^{2k+1} +\bru ^{2k+1} ) 
\sum_{h=0}^{h_0} c_k^{(h)} E_{k+1/2-2 h}(\tau )\ ,
}
$$
c^{(0)}_k={\zeta(2k+1)\ov 2k+1}\ ,\ \ \ \ h_0=[\ha (k-1)]\ .
$$
It is worth noting that the sum over $h$ contains a finite number of terms.
Terms with $h<0$ are excluded because $\aex $ must reduce to the usual genus zero result at $g_B\to 0$; terms with $h>h_0$ are related to the other terms
by the functional relation $E_r={\rm const.} E_{1-r}$. 
For $k$ odd, a given power of $s$ in eq.~\amex\ has the structure
\eqn\ccz{
g_B^{k+1/2} s^{2k+1}\big[ c^{(0)}_k E_{k+1/2}(\tau )+c^{(1)}_k 
E_{k-3/2}(\tau )+...+c^{(h_0)}_k E_{3/2}(\tau )\big] 
}
\eqn\ccv{
s^{2k+1}\big[ 
c^{(0)}_k (1+\gamma_0 g_B^{2k})+ c^{(1)}_k (g^2_B+\gamma_1 g_B^{2k-2})+...
+c^{(h_0)}_k (g^{k-1}_B+\gamma_{h_0} g_B^{k+1} )\big] 
}
whereas for $k$ even the last term in the above equations are respectively
$c^{(h_0)}_k E_{5/2}$, $c^{(h_0)}_k (g^{k-2}_B+\gamma_{h_0} g_B^{k+2})$.
Thus we have $A_4=A_4^{sl(2)}+A_{4(h=1)}^{sl(2)}+...$~,
$$
h=0\ :\ \ \ A_4^{sl(2)}\ \ \ \sim \ 1+g^2_B+...+O(e^{-1/g_B})\  \ ,
$$
$$
\ h=1\ :\ \ \ A_{4(h=1)}^{sl(2)}\ \sim \ g_B^2+g_B^4+... 
+O(e^{-1/g_B}) \ , \ \ {\rm etc.}
$$
These expressions uncover an important property of the sum over $h$: 
it adds new corrections by preserving \slz invariance and without affecting 
the leading term of the previous order. By S-duality,
the same property holds in an expansion at large $g_B$.

Defining $m=k-2h $, and in terms of Einstein-frame variables, eq.~\amex\   takes the form ($\a'=4$)
\eqn\amez{
\log \aex =-\log stu + 2 \sum_{m=1}^\infty
E_{m+1/2}(\tau ) s^{2m+1}_E \sum_{h=0}^\infty
c_m^{(h)}s^{4h}_E\ + \ (s\to t, u )\ . 
}
In string-frame variables the sum over $h$ is
$  c_m^{(0)} + c_m^{(1)} g_B^2 s^4 +
c_m^{(2)}g_B^4 s^8+... $. From eq.~\amez\ we see that
the amplitude $\aex $
generalizes $A_4^{sl(2)}(s,t)$ by replacing the constant
coefficient $c_m^{(0)}$ multiplying $E_{m+1/2}(\tau )$
by an analytic function $f_m(z)=\sum c_m^{(h)} z^h, \ 
z=\big(\a' s_E \big)^4 $
(and the same function for the terms with $t$ and $u$).
It is clear (and also implied by the theorem of sect.~5)
that the differential equation \hhh\ is not
satisfied unless $f_m $ are constants (i.e. unless 
$A_4^S(s,t)=A_4^{sl(2)}(s,t)$). 

Let us also note that if, as argued in \bevaf\ and in sect.~4, supersymmetry transformations 
relate the term $H^{4k-4}\RR^4 $ in the type IIB effective action to a term of the form $\nabla ^{4k-4} \RR^4 $, then
the constants $c_k^{(h)}$ in eq.~\amex\ are
uniquely determined to be given by $c_k^{(h)}=c_k^{(0)} \delta_{h0} $, i.e. 
$A_4^S(s,t)=A_4^{sl(2)}(s,t)$.
Indeed, an amplitude of the form
\amex\ with  $c_k^{(h)}\neq 0$ for some $h>0$ would imply, 
by supersymmetry, contributions to $H^{4k-4}\RR^4 $ of genus
lower than $k$ (cf.~\ccv ), in contradiction with the results of \bvafa . This suggests 
that generalizations of \ass\ of the form \amex\ are not possible.


To illustrate  how the  function $\aex $ may look like after resummation in $m$, we consider the following concrete example.
Let
$$
c_m^{(h)}= c_h {\zeta(2m+1)\ov 2m+1}\ ,\ \ \ \ c_0=1\ .
$$ 
Now the full series can be resummed with the result:
$$
\aex ={1\ov  s t  u}  \prod_{(p,q)'}
e^{-2\gamma (\us f(s)+\ut f(t)+\uu f(u))}
$$
\eqn\pppc{ \times\ 
\left[ { \Gamma (1- \us)\ov \Gamma (1+ \us )} \right]^{f(s)}
\left[ {\Gamma (1- \ut)\ov \Gamma (1+ \ut )} \right]^{f(t)}
\left[ {\Gamma (1- \uu)\ov \Gamma (1+ \uu )} \right]^{f(u)} \ ,
}

\md
\nn where $\gamma $ is the Euler constant, $\gamma=0.5772... $ and
$$
f(s)=1+c_1 g_B^2 \brs^4 +c_2 g_B^4 \brs^8+...
$$
The analytic structure is now more complicated.
Let us extract the complete genus one contribution
(proportional to $g_B^2$). We find
\eqn\mmbj{
A_4^{(1)}=
g_B^2 A_4^0(s,t) \left[ {2\pi^2 \ov 3} \brs\brt\bru
+ \left(-2c_1\gamma \brs^5+ c_1\brs^4 \log  { \Gamma (1- \brs)\ov \Gamma (1+ \brs )} + ( s\to t,u ) \right) \right]\ .
}
Although this exhibits the presence of cuts, it is  different
from the full genus one string amplitude (at most, it might  be a part of the  genus one string amplitud; this contains, in addition,  double poles, a non-local part coming from the massless loop contribution, etc., which
are  absent in the above expression). Note that
the first term in \mmbj\ comes from the order $h=0$, whereas
the remaining part comes from the order $h=1$.

\newsec{Eleven-dimensional origin of \ass }

So far we have examined some
properties of  \slz invariant generalizations of the
Virasoro amplitude. We will now propose an interpretation
of \ass\ in the context of superstring/M-theory.

As pointed out in sect.~2, \ass\ has poles 
corresponding to the exchange of $(p,q)$ string states.
With the exception of the  massless state $N_R=N_L=0$,
such states (which include the usual $(1,0)$ string excitations)
are expected to be unstable, since they 
do not correspond to supersymmetric classical
solutions (we recall that the states that survive in ten dimensions
have vanishing RR and NS-NS charges). 
Once all quantum corrections have been taken into account, 
poles corresponding to  unstable particles  should lie away 
from the real  axes. 
The amplitude \ass , which has no discontinuity cuts and
(for generic coupling) has simple poles lying on the real axes,
 has  the structure of a {\it tree} amplitude,
in which a certain 
collection of states are exchanged in the $s$-$t$-$u$ channels.

From the point of view of eleven dimensions  --this will be shown
 below-- 
these $(p,q)$ string states 
 of mass $\a' M^2=4N_R |p+q\tau |$ correspond to  membrane configurations
with excitations $N_R=N_L $ moving in the direction 
$(p,q)$ and carrying zero total momentum. They are not protected by supersymmetry, but these are the basic configurations that survive 
 in the ten-dimensional type IIB limit; in this limit
the BPS $(p,q)$ strings become infinitely massive and they
do not contribute to the four-graviton amplitude.
When $R_{11}\ll R_{10}$, the scattering is dominated by exchange of
$(1,0)$ strings, since all the $(p,q)$ strings with $q\neq 0$ are 
very heavy.
For $R_{11} \sim R_{10}$, the contribution of the $(1,0)$
string is of the same order as, e.g. that of the $(0,1)$ string.
It is no longer justified to
construct a perturbation theory based on the standard $(1,0)$ string,
rather than including all $(p,q)$ strings at the same time.
It must be stressed that for $g_B=R_{11}/R_{10}=O(1)$ all massive string excitations become very unstable. What this means is 
that there are other effects --in addition to the tree diagrams--
that are of the same order of magnitude (see  sect.~8).

Although in general there is not a simple 
correspondence between eleven dimensional loops and
string loops,\foot{
Further discussions on this  can be found in \refs{\ggv ,\rutse , \mgreen }.
} 
for small torus area
the correspondence is more direct.
The example of one-loop four-graviton amplitude in eleven dimensional supergravity on the torus exhibits the basic fact that
the eleven-dimensional supergravity amplitude is \slz invariant order by order in
the loop expansion (\slz symmetry being just part of reparametrization
invariance). 
But it also shows that the 1-loop contribution, when represented in terms
of the string coupling, already contains
contributions to every genus order. The reason is that at any given
loop order in eleven dimensions, additional dependence on the string coupling appears through 
the masses of the Kaluza-Klein states
running in the loops, in a way that also affects 
lower loop orders in the string perturbation theory.
Technically, this calculation applies when $R_{10},R_{11}$ are much greater
than the cutoff, given by the eleven-dimensional Planck lenght $l_p$.
In the opposite limit, $R_{10},R_{11}\ll l_{p}$,
the contribution of Kaluza-Klein states is in fact suppressed by factors
$O\big( \exp (-l_p/R)\big) $, so they do not give any perturbative contribution.

Consider, for example, the  term $\zeta(3) {\cal R}^4$ in the effective action. 
At $R_{10}R_{11}\gg l_p^2 $, it is the one-loop four-graviton amplitude in eleven dimensional supergravity that provides this contribution to the ${\cal R}^4$ term, whereas at $R_{10}R_{11}\ll l_p^2 $ such term, 
and the full Virasoro amplitude,
originate from the multiple exchange of the usual (non-BPS) string 
excitations in the tree diagram.
From the point of view   of the eleven dimensional 
theory on a torus with $R_{10}R_{11}\ll l_p^2$, 
in the Virasoro amplitude
the external gravitons are exchanging small winding membranes that 
have wave modes moving   only in one world-volume direction $\s $ (wound
around $x_{10}$). The \slz symmetry is recovered upon the inclusion of the exchange of the physical states 
representing oscillations collectively moving in an arbitrary 
$(p,q)$ direction, after summing over $(p,q)$.  
It will be argued below that (after taking $R_{10}R_{11}\to 0$) 
the \slz symmetric result that arises in this `tree-level' 
eleven-dimensional calculation is given by \ass .

\subsec{M-theory configurations corresponding to the $(p,q)$ strings}

Let us first recall the eleven-dimensional description of the
BPS $(p,q)$ strings \john .
A BPS $(p,q)$ string bound state with NS-NS and R-R charges $(lp,lq)$
with a momentum boost $w_0$ along the string becomes, after T-duality
in the string direction $x_{10}$, a bound state of a 0-brane of charge $lq$, 
a fundamental string of charge $w_0$ and a wave of momentum $lp$.
The corresponding solution in eleven dimensions was described in \rusty\ and represents an extremal 2-brane
of charge $w_0$ superposed with a gravitational wave with momentum components
$(lp,lq)$ in the directions $(x_{10},x_{11})$
(or, equivalently, momentum flux along the $(p,q)$ cycle of the torus).
In the presence of an extra translational isometry, there exists
a dual eleven-dimensional description of the same BPS $(p,q)$ string solution,
obtained by applying T-duality in the extra isometric direction $x_{9}$, and lifting to
eleven dimensions, giving a 2-brane with one leg wrapped around a $(p,q)$ cycle of the torus $(x_{9},x_{11})$, and the other leg winding $l$ 
times around $x_{10}$, superposed with a gravitational wave carrying momentum $w_0$ 
in the direction $x_{10}$ (for the explicit solution, see \trusso ).

\def\n{ {\bf n} }

At microscopic level, there are right-moving waves moving in the BPS 
$(p,q)$ string satisfying $N_R=lw_0$. There are many inequivalent physical
states with the same value of $N_R$. The corresponding
classical geometry in eleven dimensions
is the same for all of them and given by 
the fundamental membrane with the momentum boost.
We will consider a rectangular ($\tau_1=0$) torus of radii $R_{10}, R_{11}$.
The membrane coordinates can be written as follows
$$
X^{10}(\s, \rho )=w_0 R_{10}\s  + \td X^{10}(\s, \rho )\ ,\ 
$$
\eqn\mbmb{
X^{11}(\s, \rho )= R_{11} \rho + \td X^{11}(\s, \rho )\ , 
}
where $\td X^{10}, \td X^{11}$ are single-valued functions
of the membrane world-volume coordinates  $\s,\rho \in [0,2\pi)$.
The transverse membrane coordinates $X^i(\s,\rho ),\ i=1,...,8$ are all single-valued (we use the notation where 
the eleven bosonic coordinates are 
$\{ X^0, X^i, X^{10},X^{11}\} $),  
and they can be expanded in terms of
Fourier modes, 
\eqn\mbq{
X^i (\s , \rho )=\sqrt{\a'  } \sum _{k,m} X_{(k,m)}^i 
e^{i k \s+im\rho  }\ ,\ \ \ 
 P^i(\s, \rho )={1\ov (2\pi)^2 \sqrt{\a' } } \sum _{k,m}  
P^i_{(k,m)} e^{ ik \s + im\rho}\ .
}
We will assume that the dynamics of the oscillations is
governed by a relativistic membrane theory \bergsh .
Separating the winding contributions and inserting the Fourier
expansions as in ref.~\mrusso , 
the membrane light-cone Hamiltonian \refs{\bst,\dewitt }\ 
takes the form $H=H_0+H_{\rm int}$ , with \mrusso 
\def\ww{\omega_{km}}
\def\mn{ {(-k,-m)} }
$$
\a'  H_0= 8\pi^4 \a 'T_3^2 R_{10} ^2 R_{11}^2 w_0^2+  {1\ov 2} \sum _\n \big[ P_\n^a P^a_{-\n}
+\ww ^2 X^a_\n X^a_{-\n }\big]
$$
$$
\a'  H_{\rm int}= {1\ov 4g^2_A}\sum (\n_1 \times \n_2)(\n_3\times \n_4)
X_{\n_1}^a  X_{\n_2 }^b  X_{\n_3}^a   X_{\n_4}^b  
$$
$$
+{i\ov g_A}\sum mk^2 X_{(0,m)}^{10} X^i_{(k,n)}X^i_{(-k,-n-m)}\ ,
$$
$$
X^+={X^0+\td X^{11}\ov \sqrt{2} }=x^+ +\a' p^+\tau \ ,\ \ \ 
\n \equiv (k,m)\ ,\ \ \  a,b=1,...,8,10\ ,
$$ 
where $T_3$ is the membrane tension ($[T_3]=cm^{-3}$) and 
$$
\n \times\ \n '=km'-m k'\ ,\ \ \ \ \a'=\big( 4\pi ^2 R_{11} T_3\big)^{-1}\ ,
$$
$$
g^2_A\equiv {R_{11}^2\ov \a' }=4\pi^2R_{11}^3T_3\ ,\ \ \ \ \ \ 
\ww =\sqrt { k^2 + w_{0}^2 m^2 \tau_2^2 }\ ,\ \ \ \ \ \ 
\tau _2= {R_{10}\ov R_{11} }\ .
$$
Here only the bosonic modes have been written explicitly (the inclusion of
fermion modes is straightforward). The constant $g_A$ represents the type IIA
string coupling.
The  mass  operator is given by
\eqn\massa{
M^2=2p^+p^--(p^{a})^2=2 H_0+2 H_{\rm int}-(p^{a})^2\ .
}
$H_{\rm int}$ is positive definite, and any state $|\Psi \rangle $ with 
$\langle \Psi | H_{\rm int} |\Psi \rangle \neq 0$ will have 
infinite mass in the zero area limit, where $g_A\to 0$
(with $T_3\to \infty $, so that $T =2\pi R_{11} T_3$ and 
$\tau_2 $ remain fixed).
The only states that survive  are
those containing excitations in a Cartan subspace of the area-preserving
diffeomorphism algebra, so that $H_{\rm int}$ drops out from
$\langle \Psi | M^2 |\Psi \rangle $, i.e.
$$
\langle \Psi _{\rm cartan}| H_{\rm int} |\Psi _{\rm cartan}\rangle = 0\ .
$$
To be precise, le us introduce  mode operators as follows:
\def\www {w _{ {(k,m)} } }
\eqn\modoz{
X^a_{(k,m)}={i\ov \sqrt {2} \www }\big[\a^a_{(k,m)}+\td \a^a_{(-k,-m)}\big]
\ ,\ \ \ \ P^a_{(k,m)}={1\ov \sqrt {2} }\big[\a^a_{(k,m)}-\td \a^a_{(-k,-m)}\big]\ ,
}
$$
\big( X_{(k,m)}^a\big) ^\dagger =X_{(-k,-m)}^a\ ,\ \ \ \ 
\big( P_{(k,m)}^a\big) ^\dagger =P_{(-k,-m)}^a\ ,\ \ \ \ 
\www \equiv  \epsilon (k ) \ \ww  \ ,
$$
where $\epsilon (k )$ is the sign function.
The canonical commutation relations imply
$$
\big[ X^a_{(k,m)} , P^b_{(k',m')} \big]= i\delta_{k+k'}\delta_{m+m'}\delta^{ab}\ ,
$$
\eqn\zzs{
[ \a _{ {(k,m)} }^a , \a^b_{(k',m')}]= \www \delta _{k+k'}\delta _{m+m'}\delta^{ab}\ ,\  
}
and similar relations  for the $\tilde \a _{ {(k,m)} } ^a$.
Let us write $(k,m)=n(\td p,\td q)$, with $(\td p,\td q)$ relatively prime.
A {\it Cartan subspace} is constituted of all states made of operators 
$\a _{ {n(\td p,\td q)} }^a,\ \tilde \a _{ {n(\td p,\td q)} } ^a$  
with the same value of $(\td p,\td q)$. We will denote this subspace
by ${\cal H}_{\td p\td q}$. The states that have finite mass in the
zero area limit (corresponding to 
the 10D type IIB limit) live in the direct sum
of ${\cal H}_{\td p\td q}$ over $\td p,\td q$ coprime.

\def\n{ {\bf n} }

In this subspace and in this limit
the interaction term can be dropped and the world-volume
theory can be described in terms of free variables.
The solution to the membrane equations of motion is given by
\eqn\xsol{
X^a (\s, \rho, \tau )= x^a +\a'  p^a \tau + 
i \sqrt{\textstyle {\a' \ov 2}} \sum_{\n \neq (0,0)}
 w_\n \inv \big[ \a _\n^a e^{ik\s +im\rho } 
+ \td \a _\n ^a e^{-ik\s -im\rho }\big] \ e^{i w_\n \tau } \ .
}
Let the momentum components 
in the directions $X^{10}$ and $X^{11}$ be given by 
$$
p_{10}={l p\ov R_{10} }\ ,\ \ \ \ p_{11}={l q\ov R_{11} }\ .\
$$
The (nine-dimensional) mass operator takes the form
\eqn\nima{
M^2= {l^2p^2\ov R_{10}^2} + {l^2q^2\ov R_{11}^2} + {w_0^2 R_{10}^2\ov 
\a  ^{\prime 2}}\  + {2\ov \a' } {\bf H} \ ,\ \ \ \ 
}
$$
{\bf H}= \ha \sum _\n \big( \a^a_{-\n} \a^a_{\n} + \td \a^a_{-\n} \td \a^a_{\n}\big)\ ,\ \ \ \ \ \n\equiv (k,m)\ .
$$
\def\www {\omega _\n }
The level-matching conditions are given by \mrusso
$$
N_\s^+ -N_\s^-= w_0 lp\ ,\ \ \ \ \ N_\rho^+- N_\rho ^- = lq \ ,
$$
where
$$
N^+_\s = \sum _{m=-\infty }^\infty \sum _{k=1}^\infty {k\ov \ww }
\a^i_\mn \a^i_{(k,m)} \ ,
$$
$$
N^-_\s = \sum _{m=-\infty }^\infty \sum _{k=1}^\infty {k\ov \ww }
\td \a^i_\mn \td \a^i_{(k,m)} \ ,
$$
$$
N^+_\rho=\sum _{m=1}^\infty \sum _{k=0}^\infty {m\ov \ww }
\big[ \a^a_\mn \a^a_{(k,m)} + \td \a^a_{(-k,m)} \td \a^a_{(k,-m)} \big]\ ,
$$
$$
N^-_\rho=\sum _{m=1}^\infty \sum _{k=0}^\infty {m\ov \ww }
\big[ \a^a_ {(-k,m)}\a^a_{(k,-m)} + \td \a^a_{(-k,-m)} \td \a^a_{(k,m)} \big]\ .
$$
For states living in a given subspace ${\cal H}_{\td p\td q}$,
with $(k,m)=n(\td p,\td q)$, the frequency of oscillations 
becomes $w_{km}=n \sqrt{\td p^2+\td q^2 w_0^2\tau_2^2}$.
In the target space $(x_{10},x_{11})$, the corresponding
 oscillations travel in a direction 
$\theta $ 
relative to the direction $x_{10}$, with
$$
\tan\theta  = { \td q w_0\tau_2\ov \td p} \ .
$$
For the  BPS $(p,q)$ string states,
supersymmetry of the classical solution allows to add
right-moving waves only  along the momentum vector
$l (p,q)$.
This imposes a restriction on the possible states that can
microscopically describe the BPS solution.
That is, they  live in ${\cal H}_{\td p\td q}$, 
with ${\td p\ov \td q w_0}={p\ov q}$, with the extra condition that there 
are only right-moving excitations, which sets 
$\td \a ^i _{(k,m)}=0$. 
Because $(p,q)$ and $(\td p,\td q)$ are pairs of relatively prime integers,
the relation ${\td p\ov \td q w_0}={p\ov q}$ implies that one of the following two
cases is true: $(\td p,\td q)=(pw_0,q)$ or $(\td p,\td q)=(p,q/w_0)$.
To be specific, let us consider this last one (and $p\neq 0$).
The discussion for the other case is similar.
In the subspace ${\cal H}_{\td p\td q}$ the mass operator \nima\ reduces to
\eqn\mma{
M^2= {l^2 p^2\ov R_{10}^2} +{l^2 q^2\ov R_{11}^2} +{w^2_0R_{10}^2\ov {\a'}^2}+{2\ov\a '} \sqrt{p^2+q^2\tau_2^2}\ (N_R+ N_L)\ ,
}
\eqn\mmaz{
N_R-N_L=lw_0\ ,\ \ \ N_R=\sum_{n=1}^\infty \a_{-n}^i \a_{n}^i\ ,
\ \ \ \ N_L=\sum_{n=1}^\infty \td \a_{-n}^i \td \a_{n}^i\ ,
}
where the $\a_n^i $ are defined by
\eqn\hhdd{
\a_n^i= (p^2+q^2\tau_2^2)^{-1/4}\a _{(n p,n q/w_0)}^i \ ,\ \ \ \ [\a_n^i,\a_{n'}^j]=n\delta_{n+n'} \delta^{ij}\ ,
}
and similarly for $\td\a_n^i$.
Setting $\td \a ^i _{(k,m)}=0$, so that $N_L=0$, one reproduces the standard mass formula for the BPS states, as discussed in \rusty .

It is remarkable that the  mass spectrum \mma , \mmaz\
 exactly coincides with the
$(p,q)$ string mass spectrum, 
even if it includes non-supersymmetric states with $N_L\neq 0$.
The sector $l=0$ (relevant to ten dimensions)
is constituted of both right and left moving 
excitations satisfying $N_R=N_L$, and they describe
the membrane states that have finite mass in the limit
that the area goes to zero. 
Since in  eleven dimensions these are waves moving along a
$(p,q)$ direction of a membrane with winding $w_0$, which  carry  zero total momentum,
the large distance geometry 
must approximate that of a non-extremal 
black 2-brane \guven\ of charge $w_0$,
irrespective of the microscopic state, in particular, irrespective
of the $(p,q)$ orientation of the oscillations.

In the type IIB language, the winding of the membrane $w_0$ represents
the momentum of the type IIB string.
The above discussion is  invalid in the sector $w_0=0$,
that we do not know how to treat. Nevertheless, 
in the zero area limit, understanding this sector may not  be  essential:
for $R_{10}\to 0$, one can recover all continuum values of $w_0R_{10}$
(including zero) by formally starting with $w_0\neq 0$.

\subsec{Heuristic derivation of \ass }

Having argued that  
membrane theory on a vanishing torus area
is constituted by different decoupled sectors $(p,q)$,
each one being described by a free string theory 
with tension $T_{pq}=T \sqrt {p^2+q^2\tau_2^2}$,
we now examine very schematically a possible way to 
derive an $SL(2,{\bf Z})$-invariant four-graviton scattering amplitude 
starting from membrane theory.

The four-graviton scattering amplitude in string theory
is formally given by
\eqn\mema{
A_4=\langle (k_3,\zeta_3) ; (k_4,\zeta_4) 
|(k_1,\zeta_1) ; (k_2,\zeta_2) \rangle
=\int_B [DX] \ e^{S}=\int [DX] \ e^{S} \ V_1V_1 V_3 V_4 \ ,
}
where $B$ stands for the boundary carrying the information
about the quantum numbers of ingoing and outgoing states.
By conformal invariance, $B$ can be supplanted
 by the insertion of suitable vertex operators $V_i=V(k_i;\zeta_i)$.
As in the previous subsection, fermion variables are omitted.

The explicit calculation of the  path integral \mema\ gives
the Virasoro amplitude \vrz . The contribution due to massless
exchange can be obtained by taking the limit $\a'\to 0$
in \vrz . This gives
$$
A_4\bigg|_{\rm zero}=\kappa^2 K {1\ov stu } \ .
$$

From eleven-dimensional point of view, the amplitude
\mema\ is only accounting for the exchange of those membranes that
do not oscillate in the eleventh dimension.
If $R_{10}\sim R_{11}$, the contribution of membrane exchange
with oscillations in an arbitrary $(p,q)$ direction on $T^2$ will
be equally important.
The full amplitude will be given by
\eqn\mmbr{
 A_4=\langle (k_3,\zeta_3) ; (k_4,\zeta_4) 
|(k_1,\zeta_1) ; (k_2,\zeta_2) \rangle
=\int_B [DX(\s ,\rho )] \ e^{S} \ .
}
Let us now write
\eqn\unoz{
X^i (\s , \rho )=\sum_{ (p, q)'} X_{ p q}^i \ ,\ \ \ 
X_{ p q}^i=\sqrt{\a'  } \sum _{n} X_{n( p, q)}^i 
e^{i n( p \s+ q\rho)  }\ ,\ \ \ 
}
so that
$$
 [DX]=\prod_{ (p, q)'} [DX_{ p q}]\ .
$$
In the zero-area limit, the dynamics resembles that described by 
a direct sum of free-string theory lagrangians ${\cal L}_{pq}$, differing only in the tension $T_{pq}$.
The amplitude takes the form
\eqn\nono{
A_4=\prod _{ (p, q)'} \int_{\td B_{pq}} [DX_{ p q}] \ e^{S_{pq}}\sim 
\prod_{(p,q)'} \int [DX_{pq}] \ e^{S_{pq}} \ V_1V_1 V_3 V_4 \ .
}
In this way the amplitude looks like an infinite
product of string-theory amplitudes. 
There is, however, an important difference: in eq.~\nono , 
there is only one variable describing the center-of-mass coordinate (the membrane has only one center-of-mass 
mode -- $(k,m)=(0,0)$ in eq.~\mbq  ~).
A direct product of string amplitudes with independent 
center-of-mass coordinates would not give the
correct answer; in particular,  each factor would be independently accounting for  the exchange
of a string in the zero-mode state (i.e. the exchange of the massless multiplet). This would lead to the appearance of the
factor $ K {1\ov stu }$ an infinite number of times.
But it is clear from the original membrane-theory formulation that
there must be only one factor $K {1\ov stu }$, corresponding to the 
exchange of a membrane with no oscillations (representing the massless supergravity multiplet).
The $SL(2,{\bf Z})$-invariant 
amplitude schematically represented in \nono\ therefore
seems to have the structure of a product over $(p,q)'$
with a single factor $K {1\ov stu }$, just as 
the $S$-dual amplitude  given by eq.~\vvzz .
Obtaining a complete proof along these lines --i.e. starting from 
eleven dimensions and then taking the zero-torus area limit--
may be a very complicated way.
In eleven dimensions membrane theory is non-linear, and it is
only in the limit of vanishing torus area that 
the theory seems to simplify.
A more convenient approach may already exist 
in the ten-dimensional type IIB theory, without any reference to
eleven dimensions  (perhaps in the spirit of ref.~\towns ).

\newsec{Discussion}

In section 7 the amplitude \ass\ has been
interpreted as a {\it tree-level} amplitude in M-theory compactified on a 2-torus of small area.
It is a  well-known fact that in eleven dimensions there is no extra parameter that one can use to
control  loop corrections.
The model of sect.~6 may be regarded as an example on how 
other effects might be systematically organized.
The physical idea of this organization is the following.
Loop diagrams can be constructed from tree diagrams
by using unitarity (this is  the way string loops were
originally constructed).
This is not a straightforward calculation,  in particular,
one first needs to symmetrize other tree-level (N-graviton) amplitudes 
of string theory, etc.
Using  \ass\ as starting (tree-level) amplitude and assuming that
the $(p,q)$ string states
 constitute a complete set of intermediate states,
this procedure generates an \slz invariant loop expansion.
In section 7 we have presented evidence that this is 
the natural organization that follows from eleven-dimensional
 membrane theory in an expansion in topologies.
The full amplitude defined 
in this way may represent the four-graviton amplitude 
in M-theory compactified on a  2-torus in the limit
the area goes to zero at fixed modulus $\tau $.

A natural question is whether in the limit of weak (or strong)
coupling  such $SL(2,{\bf Z})$-invariant 
organization (which should follow automatically from eleven dimensions 
by virtue of reparametrization symmetry) 
represents an improvement of perturbation theory.
It is not obvious that this will be the case, since
the new contributions seem to correspond to the exchange 
of very unstable objects, and some 
of them may not even exist for a given  $g_B\ll 1$.
Although improving perturbation theory is not the aim of this work, 
 we nevertheless expect that for any coupling $g_B$
\ass\ is  
closer to the exact scattering amplitude than what the Virasoro amplitude is:
at small coupling \ass\  becomes the Virasoro 
amplitude plus additional corrections; some of them (momentum$^8$ term,
and the higher derivative terms related by supersymmetry to 
$H^{4k-4}\RR^4 $) 
are believed to contain the exact function of the coupling, and the remaining ones have the correct form to be interpreted as genuine perturbative
and non-perturbative contributions in superstring theory.

The symmetrization based on $(p,q)$ `strings'
may as well be regarded as a trick to incorporate D-instantons in the 
perturbative series (in addition to incorporating certain parts of higher genus perturbative corrections). 
Mathematically, one is summing over all possible \slz\ rotations
of the tree-level expressions. Physically, we seem to be 
accounting for the
exchange of all possible membrane states surviving in the
zero area limit (not just those with oscillations along $x_{10}$).

\bs\bs

\noindent{\bf Acknowledgements}
\medskip
I would like 
to  acknowledge the support  of the European
Commission TMR programme grant  ERBFMBI-CT96-0982.

\listrefs

\vfill\eject\end

\bigskip
\bigskip

\noindent{\bf Acknowledgements}
\medskip
I would like to thank  M. Green and A. Tseytlin for useful remarks.
I also  wish to thank SISSA for  hospitality, and  
to  acknowledge the support  of the European
Commission TMR programme grant  ERBFMBI-CT96-0982.
\vfill\eject
\listrefs
\end